\documentclass[12pt,letterpaper]{article}

\usepackage{amsmath}
\usepackage{amsfonts}
\usepackage{amssymb}
\usepackage{graphicx}
\usepackage{hyperref}

\DeclareRobustCommand{\SkipTocEntry}[4]{}

\DeclareFontFamily{OT1}{rsfs10}{}
\DeclareFontShape{OT1}{rsfs10}{m}{n}{ <-> rsfs10 }{}
\DeclareMathAlphabet{\mathscript}{OT1}{rsfs10}{m}{n}

\def\be{\begin{equation}}
\def\ee{\end{equation}}
\def\ba{\begin{eqnarray}}
\def\ea{\end{eqnarray}}

\begin{document}
\begin{center}
{\Large{\it Foreword: Advances in Astronomy Special Issue on\\[0.1cm] 
{\bf Testing the Gaussianity and Statistical Isotropy\\[0.2cm] of the Universe}}}\\[0.4cm]
{\large Guest Editors: Dragan Huterer\footnote{Department of
    Physics, University of Michigan, 450 Church St, Ann Arbor, MI 48109-1040,
    USA}, Eiichiro Komatsu\footnote{Texas Cosmology Center and Department of Astronomy, The University of Texas at Austin, Austin, Texas 78712, USA}, Sarah
  Shandera\footnote{Perimeter Institute for Theoretical Physics, Waterloo,
    Ontario, Canada}}
\end{center}

%% ABSTRACT: The last few years have seen a surge in excitement about
%% measurements of statistics of the primordial fluctuations beyond the power
%% spectrum. New ideas for precision tests of Gaussianity and statistical
%% isotropy in the data are developing simultaneously with proposals for a wide
%% range of new theoretical possibilities. From both the observations and theory,
%% it has become clear that there is a huge discovery potential from upcoming
%% measurements.  In this Special Issue of Advances in Astronomy we have
%% collected articles that summarize the theoretical predictions for departures
%% from Gaussianity or statistical isotropy from a variety of potential sources,
%% together with the observational approaches to test these properties using the
%% CMB or large-scale structure. We hope this collection provides an accessible
%% entry point to these topics as they currently stand, indicating what direction
%% future developments may take and demonstrating why these questions are so
%% compelling. The Special Issue is available at
%% \href{http://www.hindawi.com/journals/aa/2010/si.gsiu.html}{http://www.hindawi.com/journals/aa/2010/si.gsiu.html},
%% and individual articles are also available on the arXiv.

\section*{\large  Background and Motivation}

The last few years have seen a surge in excitement about measurements of
statistics of the primordial fluctuations beyond the power spectrum. New ideas
for precision tests of Gaussianity and statistical isotropy in the data are
developing simultaneously with proposals for a wide range of new theoretical
possibilities. From both the observations and theory, it has become clear that
there is a huge discovery potential from upcoming measurements.

The twin principles of statistical isotropy and homogeneity are a crucial
ingredient in obtaining most important results in modern cosmology. For
example, with these assumptions temperature and density fluctuations in
different directions on the sky can be averaged out, leading to accurate
constraints on cosmological parameters that we have today.  Nevertheless,
there is no fundamental reason why these must be obeyed by our universe.
Statistical isotropy and homogeneity are starting to be sharply tested using
the Cosmic Microwave Background (CMB) and large-scale structure
data. Recently, there has been particular activity in these areas, given
Wilkinson Microwave Anisotropy Probe's remarkable maps, combined with claims
of large-angle 'anomalies' indicating departures from statistical isotropy as
predicted by standard inflationary models.

The statement that primordial curvature fluctuations are nearly Gaussian on
scales measured by the CMB is remarkably precise, but
doesn't reveal much about their source. Current constraints on the amplitude
of the three-point correlation function of fluctuations are nearly four orders
of magnitude above predictions from single field slow-roll inflation models
and at least an order of magnitude above what is expected just from
non-linearities that develop after the primordial spectrum is laid down. There
is a wide spectrum of interesting models that can be ruled out by tightening
this constraint; conversely, a detection of non-Gaussianity would rule out
single field slow-roll inflation.  While current observations of the CMB
fluctuations provide reasonably strong evidence for a primordial source of
fluctuations from inflaton, only measurements of higher order statistics can
truly shed light on the physics of inflation.

Departures from statistical isotropy and Gaussianity involve a rich set of
observable quantities, with diverse signatures that can be measured in the CMB
or in large-scale structure using sophisticated statistical methods. These
signatures, which carry information about physical processes on cosmological
scales, have power to reveal detailed properties of the physics responsible
for generating the primordial fluctuations. Even qualitative observational
features can identify key properties of the fields involved (for example, how
many fields and which couplings were most relevant), or alternatively, shed
light on the systematic errors in the data. However, because there are so many
possibilities from both theory and observation, and because many calculations
are very technical involving methods such as higher order perturbation theory,
the literature can be daunting.

In this Special Issue of Advances in Astronomy we have collected articles that
summarize the theoretical predictions for departures from Gaussianity or
statistical isotropy from a variety of potential sources, together with the
observational approaches to test these properties using the CMB or large-scale
structure. We hope this collection provides an accessible entry point to these
topics as they currently stand, indicating what direction future developments
may take and demonstrating why these questions are so compelling. The Special
Issue is available at
\href{http://www.hindawi.com/journals/aa/2010/si.gsiu.html}{http://www.hindawi.com/journals/aa/2010/si.gsiu.html},
and individual articles are also available on the arXiv.

\section*{\large Table of Contents: Invited Reviews}
\begin{itemize}
\item {\it Non-Gaussianity from Large-Scale Structure Surveys}
  (\href{http://arxiv.org/abs/1001.5217}{arXiv:1001.5217})\\ Licia Verde
\item {\it Non-Gaussianity and Statistical Anisotropy from Vector Field
  Populated Inflationary Models}
  (\href{http://arxiv.org/abs/1001.4049}{arXiv:1001.4049})\\ Emanuela
  Dimastrogiovanni, Nicola Bartolo, Sabino Matarrese, Antonio Riotto
\item {\it Cosmic Strings and Their Induced Non-Gaussianities in the Cosmic
  Microwave Background}
  (\href{http://arxiv.org/abs/1005.4842}{arXiv:1005.4842})\\ Christophe
  Ringeval
\item {\it Ekpyrotic Nongaussianity: A Review}
  (\href{http://arxiv.org/abs/1001.3125}{arXiv:1001.3125})\\ Jean-Luc Lehners
\item {\it Primordial Non-Gaussianity in the Large-Scale Structure of the
  Universe} (\href{http://arxiv.org/abs/1006.4763}{arXiv:1006.4763})\\ Vincent
  Desjacques, Uro$\check{s}$ Seljak
\item {\it Primordial Non-Gaussianity and Bispectrum measurements in the
  Cosmic Microwave Background and Large-Scale Structure}
  (\href{http://arxiv.org/abs/1001.4707}{arXiv:1001.4707})\\ Michele Liguori,
  Emiliano Sefusatti, James R. Fergusson, E.P.S. Shellard
\item {\it Testing Gaussianity, Homogeneity, and Isotropy with the Cosmic
  Microwave Background}
  (\href{http://arxiv.org/abs/1002.3173}{arXiv:1002.3173})\\ L. Raul Abramo,
  Thiago S. Pereira
\item {\it Primordial Non-Gaussianity in the Cosmic Microwave Background}
  (\href{http://arxiv.org/abs/1006.0275}{arXiv:1006.0275})\\ Amit P.S. Yadav,
  Benjamin D. Wandelt
\item {\it Primordial Non-Gaussianities from Inflation Models}
  (\href{http://arxiv.org/abs/1002.1416}{arXiv:1002.1416})\\ Xingang Chen
\item {\it Non-Gaussianity from Particle Production during Inflation}
  (\href{http://arxiv.org/abs/1010.5507}{arXiv:1010.5507})\\ Neil Barnaby
\item {\it Review of Local Non-Gaussianity from Multifield Inflation}
  (\href{http://arxiv.org/abs/1002.3110}{arXiv:1002.3110})\\ Christian
  T. Byrnes, Ki-Young Choi
\item {\it Non-Gaussianity and the Cosmic Microwave Background Anisotropies}
  (\href{http://arxiv.org/abs/1001.3957}{arXiv:1001.3957})\\ Nicola Bartolo,
  Sabino Matarrese, Antonio Riotto
\item {\it Second-Order Gauge-Invariant Cosmological Perturbation Theory:
  Current Status}
  (\href{http://arxiv.org/abs/1001.2621}{arXiv:1001.2621})\\ Kouji Nakamura
\item {\it Large-Angle Anomalies in the CMB}
  (\href{http://arxiv.org/abs/1004.5602}{arXiv:1004.5602})\\ Craig J. Copi,
  Dragan Huterer, Dominik J. Schwarz, Glenn D. Starkman
\item {\it A comprehensive overview of the Cold Spot}
(\href{http://arxiv.org/abs/1008.3051}{arXiv:1008.3051})\\ Patricio Vielva
\end{itemize}

\end{document}